# Gate tunable parallel double quantum dot in InAs double-nanowire junctions

S. Baba[1,a], S. Matsuo[1], H. Kamata[2], R. S. Deacon[3], A.Oiwa[4], K. Li[5], H. Q. Xu[5], and S. Tarucha[1,2]

*1 Department of Applied Physics, University of Tokyo, Bunkyo-ku, Tokyo, Japan*

*2 Center for Emergent Materials Science, RIKEN, Wako, Saitama, Japan*

*3 Chief Scientist Laboratories RIKEN, Wako, Saitama, Japan*

*4 The Institute of Scientific and Industrial Research, Osaka University, Ibaraki, Osaka, Japan*

*5 Key Lab for the Physics and Chemistry of Nanodevices, Peking University, China*

## ABSTRACT

We report fabrication and measurement of a device where closely-placed two parallel InAs nanowires (NWs) are contacted by source and drain normal metal electrodes. Established technique includes selective deposition of double nanowires onto a previously defined gate region. By tuning the junction with the finger bottom gates, we confirmed the formation of parallel double quantum dots, one in each NW, with a finite electrostatic coupling between each other. With the fabrication technique established in this study, devices proposed for more advanced experiments, such as Cooper-pair splitting and the observation of parafermions, can be realized.

______________________

a) Electronic mail: baba@meso.t.u-tokyo.ac.jp

**TEXT**

Inherently low-dimensional semiconductors or bottom-up nanostructures such as nanowires, nanotubes, and self-assembled quantum dots have been attractive research materials, because they are relatively easily (or with a smaller number of gate electrodes) fabricated into effective 0-dimentional or 1-dimentional devices compared to other electron systems such as a 2-dimensional electron gas (2DEG), and provide an experimental platform for investigating the effects of quantum confinement and interactions in electronic systems. Among such materials, InAs nanowires (InAs NWs) have been particularly widely studied. The initial studies on their electric properties revealed the formation of quantum dots (QDs), and their gate tunability[1–3], which later led to the experiment of controlling electron spins as qubits with double quantum dots made in single InAs NWs [4–7], taking advantage of the large g-factor[4] and spin-orbit interaction[8] of InAs. Other efforts have been made in systems where InAs NWs are contacted to superconducting electrodes. Because InAs NWs can form an ideal interface with superconducting metals without Schottky barriers, non-dissipative supercurrent transistors have been fabricated[9]. More advanced experiments include studies on Cooper-pair splitters[10–12] and realization of Majorana fermions[13,14], both of which are based on the spin correlation of Cooper pairs in superconducting electrodes.

The bottom-up nanostructures, however, have some difficulties in fabricating into complex devices like coupled NWs. Major difficulties are positioning, making contact electrodes, and surface cleaning. Indeed very few studies have been reported on the fabrication and characterization of multiple NW devices. Then the first question is whether one can fabricate a system with multiple NWs in proximity so that there is a finite and tunable coupling between QDs. In this study we repot on fabrication and characterization of InAs DNW junctions, where a quantum dot is formed in each NW, and evaluated the gate performances. The DNW devices are useful for making multiple spin qubits and topological circuits of non-local entangled electrons[15–19] or parafermions as well as majorana fermions[20]. We develop a fabrication technique for selectively placing DNW onto previously fabricated gate arrays and selectively contacting the two NWs. We characterize the gate performance of the InAs NW devices to control the electronic states of the two wires independently.

The nanowires used in this study is grown by CBE (Chemical Beam Epitaxy). Pure gold particles obtained through aerosol deposition on the substrate were used as seeds for nanowire vapor-liquid-solid mechanism growth. The length of nanowires ranges from 2 µm to 4 µm, the diameter from 60nm to 80nm. By Transmission Electron Microscopy it is confirmed that the nanowires have a pure wurtzite structure and all grow perpendicularly to the substrate surface, along the <111> direction.

Our DNW junctions were fabricated in the following way. On a $SiO_2$/Si substrate we first fabricate arrays of Ti/Au (5nm/20nm) finger bottom gates with conventional electron beam lithography. Their width and center-to-center distance are 50nm and 120 nm, respectively. Next, an insulating silicon nitride layer (35nm) is grown with CVD (Chemical Vapor Deposition), in order to electrically isolate the gate array from the junction which is later fabricated.

Then, in transferring nanowires, we employ a specific technique for selectively place multiple nanowires onto the gate regions defined above. First we pick up innumerable nanowires with a cotton bud from a growth substrate, then transfer them to an "intermediate substrate" which is covered with a PVA film and a PMMA film (See Fig. 1(a) (i) and (ii)). At this point, with an optical microscope, it is possible to locate two nanowires that are stuck together (Fig. 1 (b)), which actually have brighter contrast than single nanowires. Use of SEM is avoided due to the possible damage – or excessive injection of electrons – to the nanowires. With a proper treatment including the heating of the substrate, the PMMA layer can be peeled off. The film is then flipped, and attached to a home-made micro manipulator.

Next, the film is contacted onto the device substrate so that the target nanowires will be placed on a gate array (Fig. 1(a) (iii)). For this precise alignment, we observe the film and the substrate with the optical microscope, and carefully align them with the manipulator. In order to avoid placing unnecessary nanowires to the device substrate, we cover the substrate with PMMA and open a window over the gate array with electron beam lithography.

After NWs are transferred, the device substrate is heated up to 180℃, to enhance the adhesion of the NWs. Then, the PMMA film used for the transfer is chemically removed with accetone. Figure 1 (c) shows the NWs and gate arrays at this point. Finally we deposit Ti/Au (5nm/125nm) contacts to the NWs. Before the contact deposition the nanowires were etched in the $(NH_4)_2S_x$ solution, in order to remove the native surface oxide layer of InAs nanowires. We also fabricated sidegates to tune the NWs, but in this study only the bottom gates are utilized. Fig. 2 (a) shows one of the completed devices with a junction length $L \sim 250$ nm.

In this study the nanowires are placed approximately in parallel to the bottom gates, which are designed to tune each nanowire individually. We note that it is also possible to place nanowires perpendicularly to the bottom gates, and use them to tune the lead-nanowire tunnel couplings, and tune each nanowire's electrochemical potential with sidegates.

We measured the differential conductance of the junction with the conventional lock-in technique, with the excitation voltage $V_{ex}$=100 $\mu$ V, and frequency $f_{ex}$=37Hz. All measurements were performed in the $^3$He-$^4$He dilution refrigerator, at T~50mK.

Figure 3(a) shows differential conductance of the junction with the length L ~ 150nm, measured as a function of two different finger bottom gates. Observed resonance peaks with two different slopes suggest that there is a parallel double quantum dot (QD1, QD2) formed in the junction. In addition, the absence of a background conductance supports that each quantum dot is formed on different nanowires. Although we don't have the direct control of the tunnel couplings, the potential barriers formed at the contact edge lead to the confinement of the QDs. Therefore, width of the peaks remains wide, yielding the tunnel coupling between QDs and contacts $\Gamma$~1-2meV for both NWs.

Figure 3 (b) and (c) show a differential conductance as a function of source-drain bias voltage $V_{sd}$ and the gate voltage applied to each bottom gates. In some gate region, (for example V =-5.0 V in Fig.2 (b),) clear Coulomb diamonds are observed. From the size of Coulomb diamonds, we obtained the addition energy of QD1 and QD2, ranging in the investigated region $E_{add}1$=4-5 meV and $E_{add}2$=5-7 meV. We also calculated the gate capacitances between the finger gates and QDs. From fig. 3(b)(c) Cg1-QD1 and Cg2-QD2 are obtained, while Fig3 (a) yields the relative lever arm or the two gates. Obtained capacitances are (Cg1-QD1, Cg1-QD2)=(0.5aF, 0.25aF), and for bottom gate2 (Cg2-QD1, Cg2-QD2)=(0.05aF, 0.08aF). These capacitances are much lower than the expected values ~10aF calculated from the geometry of the device and dielectric constant of SiN 7.5. We attribute this deviation to the field screening by the contact electrodes and unused sidegates. For a better gate efficiency, it would be preferable to make the junction length L longer, and define tunnel barriers of QDs with bottom gates which are perpendicular to NWs, while tuning the potential of each QD with sidegates.

We further tuned the bottom gates, and brought QDs to the region where we confirmed features of proximity of the two QDs. In Fig. 4 (a) and (b), honey-comb like features are observed at the crossings of the resonance peaks resulting from the electrostatic coupling of the two QDs, which ensures that two QDs, one in each InAs NW, are located close enough to affect each other's electrostatic potential. Obtained inter-dot coupling is $E_{cm} \sim 0.4\text{-}0.6$ meV. $E_{cm}$ is the change in the energy of one dot when an electron is added to the other dot, following the definition presented in the previous works[21]. The small distance between two nanowires is a key factor for many advanced experiments proposed for DNW systems. In contrast, we didn't observe any direct evidence of tunnel coupling between the two QDs, which would manifest itself as avoided crossings of the peaks. We attribute the absence of the tunnel coupling to the surface oxide layers of InAs NWs, which are etched only beneath the contacts. Removing the oxide layer in QD regions, and placing NWs so that they touch each other might enable the observation of the tunnel coupling. It is also possible to reduce the lead-QD tunnel couplings ($\Gamma_{\text{lead-QD}}$) with sidagates, or so that inter-dot coupling exceeds them, or to control the inter-dot tunneling itself by a finger bottom gate. Observation and control of inter-dot tunneling will widen the possibility of DNW devices as an experimental platform of quantum bit operations.

In summary, we fabricated a device where closely-placed two parallel InAs NWs are contacted by source and drain electrodes. By tuning the junction with the finger bottom gates, we confirmed the formation of parallel double quantum dots, one in each NW, with a finite electrostatic coupling between each other. To our knowledge it is the first work to report. With the fabrication technique established in this study, devices proposed for more advanced experiments, such as the observation of Majorana and Para Fermions, can be realized.

## ACKNOWLEDGEMENTS

Part of this work was supported by Grant-in-Aid for Scientific Research(S) (No. 26220710), Grant-in-Aid Research A from Nos. 25246004 and 25246005 and Innovative areas (No. 26103004), Grants-in-Aid from JSPS (Nos. 25600013, and 26706002) MEXT, Grant-in-Aid for Scientific Research on Innovative Areas "Science of hybrid quantum systems" (No. 2703), Project for Developing Innovation Systems of MEXT, FIRST program, ImPACT Program of Council for Science, Technology and Innovation (Cabinet Office, Government of Japan), MEXT Project for Developing Innovation Systems, and Japan Society for the Promotion of Science through Program for Leading Graduate Schools (MERIT). S.B. acknowledges support from Grants-in-Aid for JSPS fellows.

# FIGURES:

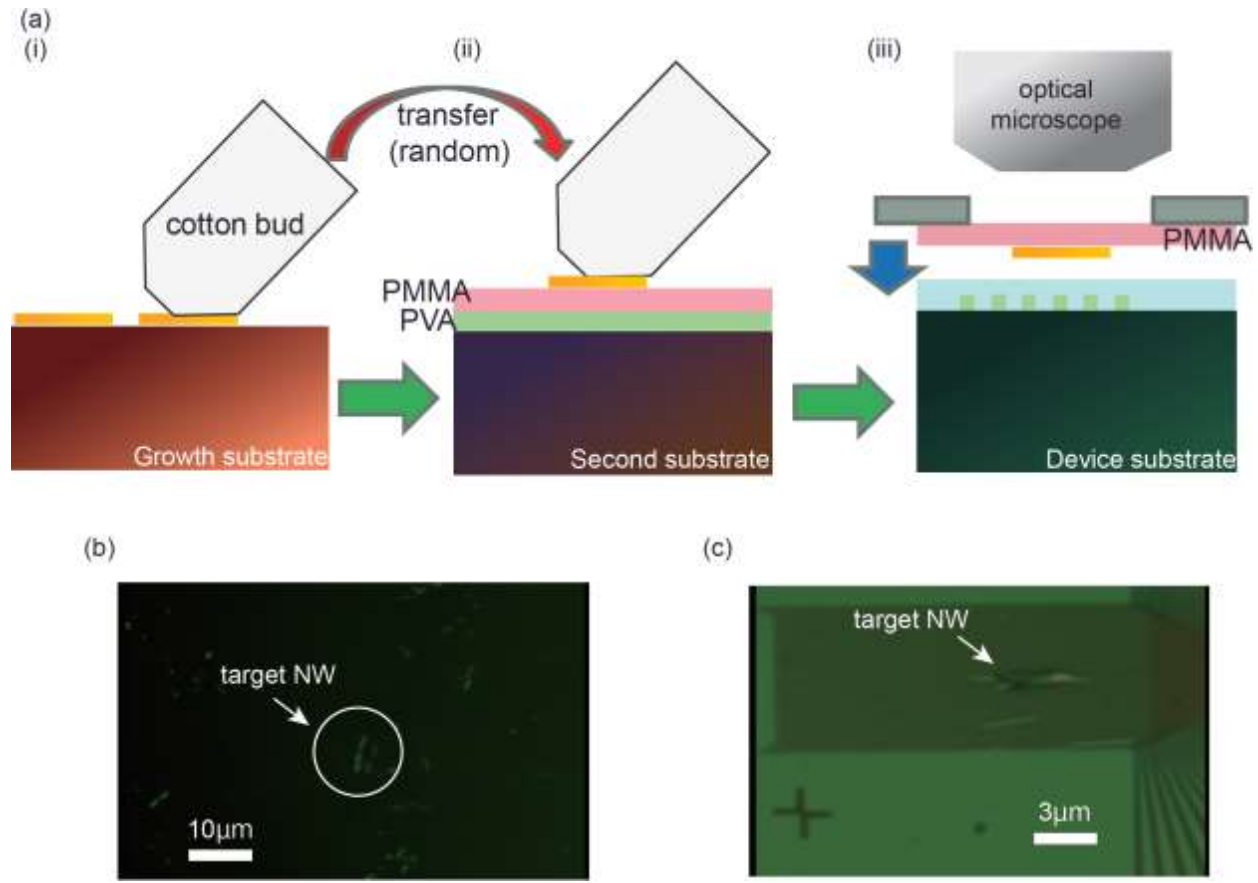

**Figure 1** (a) Illustration of the NW transfer process. (i) Many NWs are transferred from a growth substrate with a cotton bad. (ii) NWs are transferred to a second substrate which is covered with two polymer films (PVA, PMMA). (iii) The PMMA layer is peeled off and attached to a home-mad micro manipulator, which enables a fine control of the location of NWs with the observation through an optical microscope. (b) An optical micrograph of the NWs on the second substrate. Multiple NWs stuck together have a stronger contrast than single NWs, (c) An optical micrograph of the transferred NWs on the device substrate.

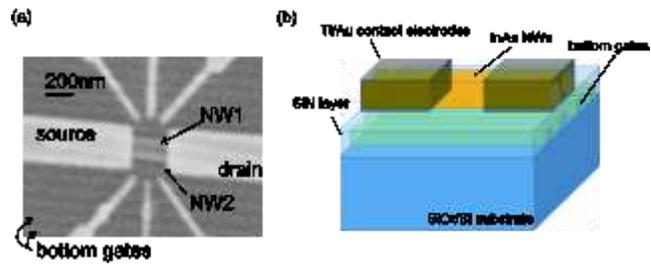

**Figure 2** (a) An SEM image of a device. Two InAs nanowires (NW1, NW2) are placed in parallel to each other, and contacted by source and drain electrodes. Beneath a Silicon Nitride insulating layer, finger bottom gates are fabricated approximately parallel to the NWs. Thin electrodes surrounding the junction are sidegates, which are not utilized in this study. (b) A schematic diagram of the device and substrate. SiN layer electrically insulates the bottom gates from the junction.

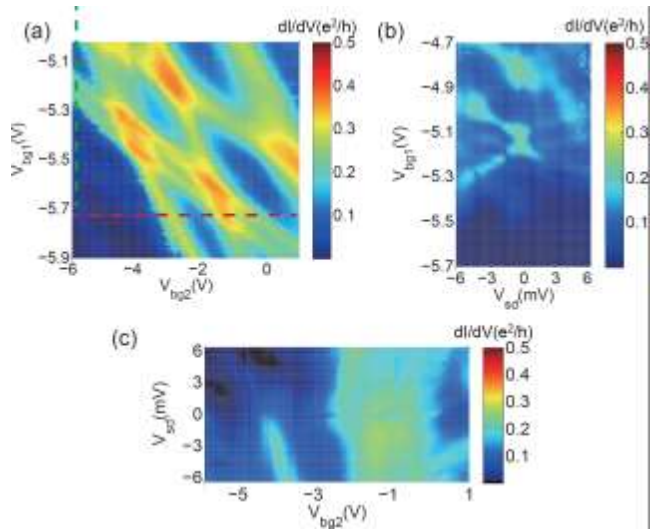

**Figure 3** (a)Charge stability diagram as a function of voltages applied two finger bottom gates $V_{bg1}$ and $V_{bg2}$. (b) Differential conductance plotted as a function of $V_{bg1}$ and $V_{SD}$, in the gate region shown in the green dashed line in (a). (c)Differential conductance plotted as a function of $V_{bg2}$ and $V_{SD}$, in the gate region shown in the red dashed line in (a).

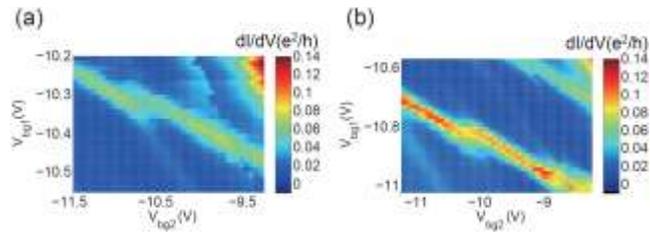

**Figure 4** (a)(b) Charge stability diagram of the regions where a honey-comb structure due to finite capacitive coupling is observed.